\documentclass[preprint,trackchanges]{aastex62}

\newcommand{\speed}[1]{#1 km~s${}^{-1}$}
\newcommand{\accel}[1]{#1 m~s${}^{-2}$}
\newcommand{\nfig}[1]{Figure~\ref{#1}}
\shorttitle{Twin CME}
\shortauthors{Duan et al.}

\begin{document}

\title{The Birth of a Jet-driven Twin CME and Its Deflection from Remote  Magnetic Fields}

\correspondingauthor{Yuandeng Shen, and Hongfei Liang}
\email{ydshen@ynao.ac.cn, lhf@ynao.ac.cn}

\author{Yadan Duan}
\affil{Yunnan Normal University, Department of Physics, Kunming 650500, Yunnan, China}
\author[0000-0001-9493-4418]{Yuandeng Shen}
\affil{Yunnan Observatories, Chinese Academy of Sciences, Kunming, 650216, China}
\affil{Center for Astronomical Mega-Science, Chinese Academy of Sciences, Beijing, 100012, China}
\author{Hechao Chen}
\affil{Yunnan Observatories, Chinese Academy of Sciences, Kunming, 650216, China}
\affil{Center for Astronomical Mega-Science, Chinese Academy of Sciences, Beijing, 100012, China}
\affil{University of Chinese Academy of Sciences, 19A Yuquan Road, Shijingshan District, Beijing 100049, China}
\author{Hongfei Liang}
\affil{Yunnan Normal University, Department of Physics, Kunming 650500, Yunnan, China}

\begin{abstract}
We report the formation of a complicated coronal mass ejection (CME) on 2015 August 23 by using the high temporal and high spatial resolution multi-wavelength observations taken by the {\sl Solar Dynamic Observatory} and the {\sl Solar and Heliospheric Observatory}. The CME exhibited both jet-like and bubble-like components simultaneously, and therefore we call it a twin CME. Detailed imaging and kinematic analysis results indicate that the twin CME were evolved form the eruption of a mini-filament driven blowout jet at the east edge of an equatorial coronal hole, in which the activation of the mini-filament was tightly associated with the continuous flux cancellation and quasi-periodic jet-like activities in the filament channel. Due to the magnetic reconnection between the filament and the ambient open field lines, the filament broke partially at the northern part and resulted in an intriguing blowout jet in the south direction. It is interesting that the ejecting jet was deflected by a group of remote open field lines, which resulted in the significant direction change of the jet from southward to eastward. Based on the close temporal and spatial relationships among the jet, filament eruption, and the twin CME, we conclude that the jet-like CME should be the coronal extension of the jet plasma, while the bubble-like one should be originated from the eruption of the mini-filament confined by the closed magnetic fields at the jet-base.
\end{abstract}


\section{Introduction} \label{sec:intro}
Solar eruptions are eruptive transient magnetic activities in the solar corona, in which magnetic energy can be released to heat and eject the magnetized plasma. Such ejection of plasma from lower to upper layers in the solar atmosphere can be divided into two classes \citep{2018ARep...62..359F}. The first class of solar eruptions is the most spectacular large-scale magnetized plasma eruption, i.e., coronal mass ejections (CMEs). Typically, they often occur accompanying with solar flares, and demonstrate as bubble-like or loop-like shape in the outer corona, and they are generally considered to be driven by magnetic structures such as magnetic flux ropes and filaments \citep[e.g.,][]{2000JGR...105.2375L,2010ApJ...725L..84L,2011RAA....11..594S,2012ApJ...750...12S,2018ApJ...853....1S,2019ApJ...874...96Y}. The projection speed of CMEs ranges from ${\sim}$ 20 km s$^{-1}$ to 2000 km s$^{-1}$, and their average speed increase from 300 km s$^{-1}$ near solar minimum to 500 km s$^{-1}$ near solar maximum \citep{2004ASPC..325..401Y}. Generally, the eruption direction of CMEs is often along the radial direction from the solar center propagating into the interplanetary space. However, some CMEs also show direction deviation with respect to the solar radial direction \citep{1986JGR....91...31M,2007ApJ...667L.105J,2008ApJ...677..699J}. The direction deviation of a CME might take place in both the latitude and longitude in the interplanetary \citep{2010NatCo...1E..74B,2013ApJ...777..167D,2013JGRA..118.6007Z,2010ApJ...722.1762L,2013SoPh..284..203I}. \citet{2004JGRA..10912105G,2009JGRA..114.0A22G} presented that corona holes are responsible for the deflection of CMEs away from the Sun-Earth line. \citet{2009AnGeo..27.4491K} suggest that  during solar minimum period, the low detection rate of interplanetary CMEs at ecliptic plane may be increased if CMEs are guided towards the equator from their high-latitude source regions by the magnetic fields in the polar coronal holes. \cite{2018ApJ...862...86Y} reported the nearly 90$^\circ$ direction change of a CME compare to the initial eruption direction, due to the deflection of the open coronal hole magnetic fields. \citet{2011SoPh..269..389S} and \citet{2011SoPh..271..111G} considered the CME deflection is related to gradients in the magnetic energy density of the background solar corona. Deflection effects of CMEs can also caused by nearby strong magnetic structures \citep[e.g.,][]{2007ApJ...667L.105J,2011NewA...16..276B,2013ApJ...773..162B}. For space weather forecasting, better we understand the deflected trajectory of a CME through the low corona to the heliosphere, the better we can predict the its near-Earth effects \citep{2013ApJ...775....5K}.

The second class of solar eruption is in form of short-term, thin, and collimated jets from lower layers of the solar atmosphere. Solar jets are ubiquitous and can be observed in active regions, quiet-Sun regions, and corona holes \citep{1992PASJ...44L.173S,1994ApJ...431L..51S,2012RAA....12..573C,2017ApJ...841L..13C,2011RAA....11.1229Y,2013SoPh..282..147L,2014Sci...346A.315T}. According to different observing wavelengths, solar jets can be divided into H$\alpha$ surges, extreme ultraviolet (EUV) jets, and X-ray jets \citep{1973SoPh...32..139R,1996ApJ...464.1016C,1999ApJ...513L..75C,2009ApJ...707L..37L,2000A&A...361..759Z,2014A&A...567A..11Z,2018ApJ...854..174T,2018ApJ...860L..25Y,2018NatSR...8.8136L,2019ApJ...871....4H}. \citet{1995Natur.375...42Y,1996PASJ...48..353Y} performed two-dimensional magnetohydrodynamics (MHD) simulations for solar X-ray jets by using two magnetic initial configurations to generate anemone and two-sided loop jets. The ejection of the two types of jets are all caused by the magnetic reconnection between emerging and nearby coronal fields. Magnetic reconnection between emerging bipoles and their ambient open magnetic field lines is considered as the cause of most collimated solar jets (e.g., \citet{1994ApJ...431L..51S,2004ApJ...610.1136L,2011ApJ...735L..43S}). When emerging fluxes reconnect with overlying horizontal magnetic fields, two-sided loop jets are produced \citep{1995Natur.375...42Y,1998SoPh..178..173K,2013ApJ...775..132J}. Besides flux emergence, many observations have indicated that flux cancellation is also important to trigger solar jets (e.g., \citet{2012ApJ...745..164S, 2017ApJ...851...67S,2012RAA....12..300Y,2014ApJ...783...11A,2016ApJ...832L...7P,2017SoPh..292..152J}). \citet{1992PASJ...44L.173S} found that the typical size of X-ray jets ranges from 5 $\times$ 10$^{3}$ km to 4 $\times$ 10$^{5}$ km, their velocity is in the range of 30 to 300 km s$^{-1}$, and the corresponding kinetic energy is of 10$^{25}$ ${\sim}$ 10$^{28}$ erg. Interestingly, solar jets sometimes can direct or indirectly launch large-scale coronal waves \citep[e.g.,][]{2018MNRAS.480L..63S,2018ApJ...861..105S,2018ApJ...860L...8S} and CMEs \citep[e.g.,][]{1998ApJ...508..899W,2002ApJ...575..542W,2005ApJ...628.1056L,2008SoPh..249...75L,2008ApJ...677..699J,2012ApJ...745..164S}. For the generation of two-sided loop jets, recent high resolution observations indicated that other mechanisms can also trigger them \citep{2017ApJ...845...94T,2018ApJ...853L..26H,2018ApJ...861..108Z,2019ApJ...871..220S}. Especially, \cite{2018NewA...65....7T} found that two-sided loop jets in filament channel can result in the loss-of-equilibrium and eruption of large-scale filaments.

\citet{2010ApJ...720..757M} proposed that there is a dichotomy of solar jets, they reclassified solar anemone jets as standard jets and blowout jets. They found about two thirds of solar anemone jets are standard jets, while the rest of one third are blowout jets. With high-resolution observations, more and more studies confirmed the finding of blowout jets \citep{2012ApJ...745..164S,2013RAA....13..253H,2016ApJ...830...60H,2017ApJ...835...35H,2014ApJ...783...11A,2015ApJ...798L..10L,2015ApJ...814L..13L,2017ApJ...836..235L,2017ApJ...842L..20L,2018Ap&SS.363...26L,2017ApJ...844L..20Z,2019ApJ...872...87L,2018ApJ...860L..25Y,2017ApJ...834...79Z,2018ApJ...869...39M,2019ApJ...877...61M}. Compare with standard jets, the blowout jets characterized initially narrow spire that later broader; initial compact brightening that spread to the whole jet-base; the most point is a cool component ($T$ $\sim$ 10$^5$ K) visible in EUV 304 \AA\ images \citep{2010ApJ...720..757M,2015Natur.523..437S}. Several studies have found the blowout jets are often associated with the eruptive mini-filament \citep[e.g.,][]{2012ApJ...745..164S,2017ApJ...851...67S, 2012NewA...17..732Y,2014ApJ...783...11A,2016ApJ...821..100S,2015Natur.523..437S}. \citet{2015Natur.523..437S} reported 20 randomly selected X-ray jets that all 20 jets originated from mini-filament eruptions, and they further suggested that standard jets and blowout jets are basically the same phenomenon, whose morphology depends on the specific situation of mini-filament eruption. \citet{2011ApJ...735L..43S,2013RAA....13..253H} indicated that some blowout jets display untwisting motion, which may result from the erupting mini-filaments or strongly twisted structures in the jet base. These observations suggested that mini-filament eruptions play an important role in the generation of solar jets. The untwisting motion and the origin of the cool component in jet base thus have been well explained because of involving of mini-filament \citep{2011ApJ...735L..43S,2012ApJ...745..164S,2017ApJ...851...67S}.

Previous observations have suggested that small-scale solar jets have significant similarities with larger-scale solar eruptions. Investigating and revealing their similarities and differences is of great significance to explain multi-scale solar eruption phenomena. In particular, \citet{2012ApJ...745..164S} for the first time reported that a pair of simultaneous CMEs are directly evolved  from a single coronal blowout jet, in which one is in jet-like shape, the other is in bubble-like shape. Indeed, it is well documented in the literature that EUV jets present a white-light counterpart observed with coronagraphs \citep{2009SoPh..259...87N,2010AnGeo..28..687N,2010SoPh..264..365P,2012A&A...538A..34F} in the form a narrow emission, hence different from the standard CME picture.
This study provided a new thought for understanding the generation of large-scale CMEs from small-scale jet activities. Afterwards, many similar events were reported \citep{2016ApJ...823..129A,2018ApJ...869...39M,2019ApJ...877...61M,2019SoPh..294...68S}, and they are well explained by the model presented in \cite{2012ApJ...745..164S}. In addition, \cite{2013ApJ...769L..21A} performed a three-dimensional numerical simulation that reproduced the observations presented in \cite{2012ApJ...745..164S}, but they did not investigate the CMEs in the outer corona. Observing and studying such kind of events is helpful for understanding the coupling phenomenon of multi-scale eruptive solar activities. However, similar studies are still very scarce.

In this paper, we report a mini-filament driven blowout jet in the low corona, which unambiguously resulted in a twin CME in the outer corona. Compared with previous studies, the advantage of this event is that eruption source region located at E40$^{\circ}$N6$^{\circ}$, suffering from less projection effect. Thus more accurate observational information during the initiation stage of the CME can be better obtained. More interestingly, we notice that the jet underwent an obvious deflection process due to its interaction with a group of remote open coronal loops, which resulted in the ejecting direction of the jet (or CME) significantly. These observations provide new clues for our understanding of coronal jets and CMEs. The paper is structured as follows: Section 2 describes instruments and data; Section 3 describes observations and results; summary and discussion are presented in Section 4.

\section{Observation And Data Analysis} \label{sec:instr}
The blowout jet occurred on 2015 August 23 in a quiet-Sun region (E40$^{\circ}$N6$^{\circ}$) near the east edge of an equatorial coronal hole. To study the generation of the jet and the associated CMEs in detail, we use observations mainly from the Atmospheric Imaging Assembly \citep[AIA;][]{2012SoPh..275...17L} on board the {\em Solar Dynamics Observatory} \citep[{\em SDO};][]{2012SoPh..275....3P}. AIA takes images of the Sun with seven extreme ultraviolet(EUV), two UV, and one visible wavelength channels. The spatial resolution of the AIA images is of 0\arcsec.6, and the cadences for the EUV, UV, and visible channels are 12 s, 24 s, and an hour, respectively. The eruption can be observed at all AIA EUV channels, and we mainly use the 171, 304, 94 and 211\AA \ images. The Helioseismic Magnetic Imager \citep[HMI;][]{2012SoPh..275..207S} line-of-sight (LOS) magnetograms are used to investigate the corresponding photospheric magnetic field evolution, which have a cadence of 720 s and a pixel size of 0\arcsec.6. The white-light observations taken by the Large Angle and Spectrometric Coronagraph \citep[LASCO;][]{1995SoPh..162..357B}) are used to study the CME evolutions. In addition, H$\alpha$ images taken by the Global Oscillation Network Group \citep[GONG;][see \nfig{fig1}(a)--(c)]{2011SPD....42.1745H} are also used, which have a pixel size of 1\arcsec and a cadence of 1 minutes. We also utilize the Potential Field Source Surface (PFSS) \citep{2003SoPh..212..165S} model to investigate the primary three-dimensional magnetic topology around the eruption source region. In addition, the simultaneous observations of WIND/WAVES are utilized to detect the associated interplanetary radio signals. To compensate for solar rotation, all the images taken at different times are differentially rotated to a reference time of 04:15 UT.

\section{RESULTS} \label{sec:OBS}
\subsection{Activation and Initiation of the Mini-filament}
\begin{figure}[t]    
\centerline{\includegraphics[width=1\textwidth,clip=]{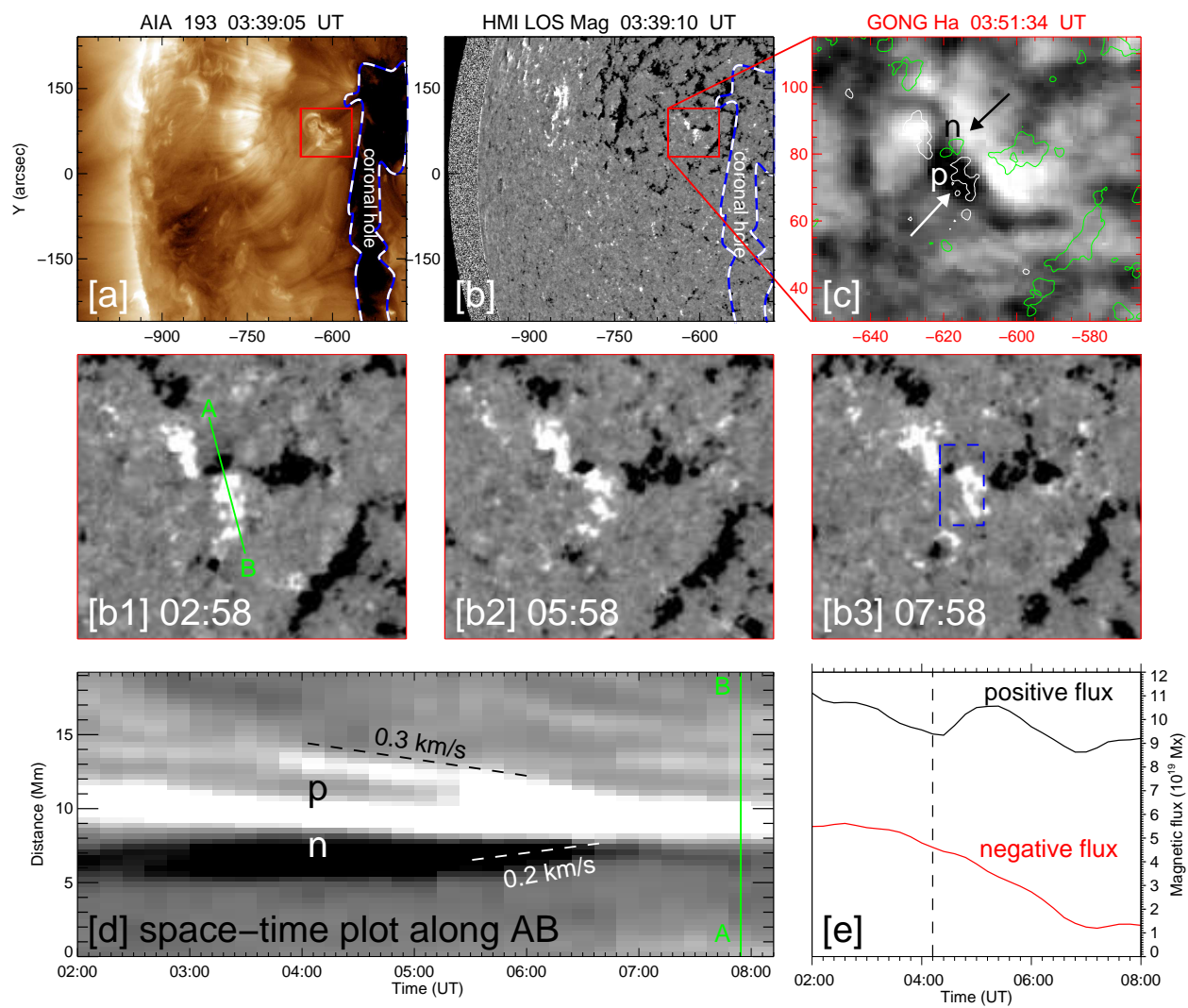}}
\caption{Panel (a) and (b) : the magnetic field environment of the event, the white-blue dashed line outlines the boundary of the equatorial coronal hole. The red boxes denote the FOV of panels (b1)--(b3) and (c). Panel (c) : a GONG H${\alpha}$ image overlaid with the  photospheric magnetic fields, in which positive and negative fields are plotted in white and green contours, and the corresponding scale levels are ${\pm}$50 G, respectively. Panels (b1)--(b3) : time sequence HMI LOS magnetograms show the flux convergence and cancellation, in which the white (black) patches are positive (negative) magnetic polarities. Panel (d): the time-distance plot along the green line A--B in (b1). Panel (e): the positive (black) and negative (red) flux variations in time in the blue dashed box in (b3) from 02:00 UT to 08:00 UT, and the black dashed line indicates the start time of the mini-filament eruption.}
\label{fig1}
\end{figure}

\begin{figure}[b]    
\centerline{\includegraphics[width=0.95\textwidth,clip=]{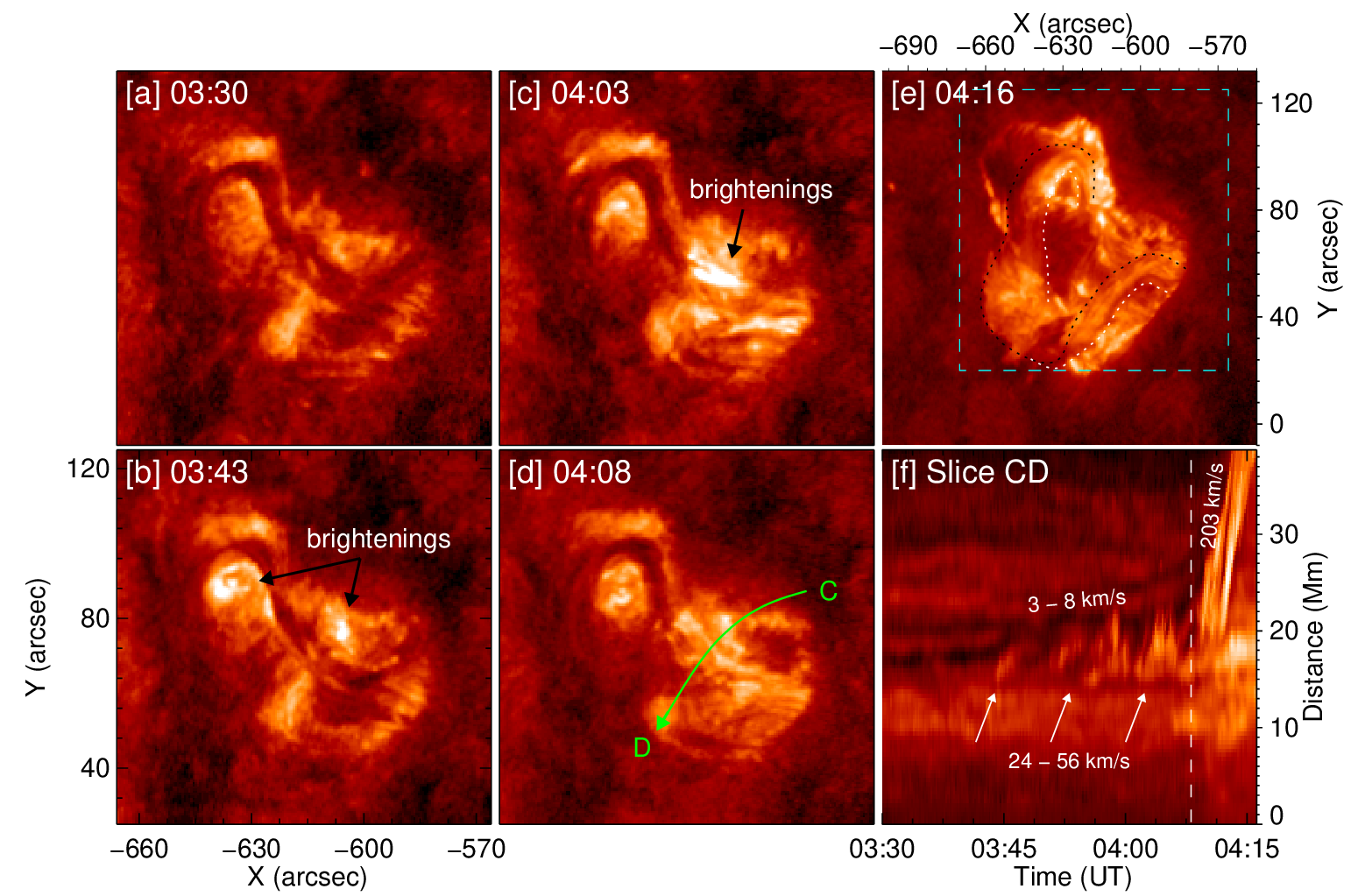}}
\caption{Panel (a)--(e): the activation and eruption of the filament observed in AIA 304 \AA\ images. The black and white dashed lines in panel (e) outlines the twisted filament, and the cyan dashed box indicates the FOV of (a)--(d). The time-distance plot along the green curve arrow C--D in panel (d) is shown in panel (f); The white arrows in (f) denote intermittent jet-like features, and the vertical white dashed line indicates the start time of the mini-filament eruption.}
\label{fig2}
\end{figure}

On 2015 August 23, a mini-filament (length $\approx 30$ Mm) resided in a mixed-polarity quiet-Sun region at the east edge of an equatorial coronal hole, and it can well be distinguished in the H$\alpha$ image (\nfig{fig1}(a--c)). In order to understand the activation and eruption of the mini-filament, we investigate the evolution of the photospheric magnetic field within the eruption source region. HMI LOS magnetograms show that continuous flux convergence and cancellation took place near the polarity inversion line (PIL) of a pair of opposite polarities. The converging motion of the opposite polarities were along the path as shown by the green line in \nfig{fig1}(b1). One can see that the area of the black negative polarity became smaller and smaller from 02:58 UT to 07:58 UT (see \nfig{fig1}(b1)--(b3)). The temporal variations of the positive and negative polarities are also showed as a time-distance plot in \nfig{fig1}(d), which better shows the convergence and cancellation processes of the opposite polarities. Here, the time distance diagram is obtained by composing the time sequence of the intensity profiles along A--B. In addition, it is also measured that the converging speeds of the positive and negative polarities were about \speed{0.3 and 0.2}, respectively. We also calculate the flux variations in the box region as shown in \nfig{fig1}(b3), and the result is shown in \nfig{fig1}(e). It is clear that the unsigned negative flux decreased from 5.0 to 1.0 $\times$ 10$^{19}$ Mx, while the positive flux decreased from 11.0 to 9.0 $\times$ 10$^{19}$ Mx. This flux changing pattern clearly suggests the flux cancellation between the positive and negative fluxes. In addition, the positive flux also showed a small bump during 04:10 UT to 06:50 UT, which indicates the emergence of the positive polarity. The start time of increasing of the positive flux was coincident with the start time of the filament eruption.

The activation and initiation of the mini-filament eruption are shown in \nfig{fig2} using the AIA 304 \AA\ images. The filament showed an inversed-S shape at about 03:30 UT (see panel (a)). A few minutes before the eruption, weak brightenings were observed at the middle section of the filament, which lasted for a few minutes and became more and more pronounced (see panel (b)--(d)). During this stage, intermittent jet-like features were observed below the filament. After the brightenings and intermittent jet-like activities, the filament started to erupt at about 04:10 UT, then it rapidly erupted to the southeast direction and accompanied with a lateral rolling motion. To further demonstrate the dynamical evolution of the filament, a time-distance plot is made along the green curve C--D in panel (d). One can see that before the filament got into its main eruption phase, intermittent jet-like features took place below the filament body, and they showed a quasi-periodic ejection behavior. The average time interval between neighboring jet-like features was about 335 seconds; they lasted for about 25 minutes from 03:45 UT to 04:10 UT and with a speed in the range of about \speed{24--56}. It is noted that before the violent eruption of the filament it underwent a slow rising phase at a projection speed of \speed{3--8} (see panel (f)).

Based on the investigation of the variation of the magnetic flux and the eruption behavior of the filament described above, it is clear that the filament activation and initiation were associated with the flux cancellation and emergence in the photosphere. Before the violent eruption stage of the mini-filament, the continuous flux cancellation caused the intermittent jet-like features and brightenings below the filament that also showed a slow rising phase of about 25 minutes. At about 04:10 UT, the sudden emergence of the positive flux can accelerate its cancellation to the negative flux; this may directly destroy the quasi-equilibrium state of the filament and therefore resulted in its violent eruption. It should be pointed out that the violent eruption of the filament could also be caused by kink instability after its activation, since strong rolling motion is observed in the erupting filament \citep{2012ApJ...750...12S}.

\subsection{Formation of the Blowout Jet}
\begin{figure}[b]    
\centerline{\includegraphics[width=0.95\textwidth,clip=]{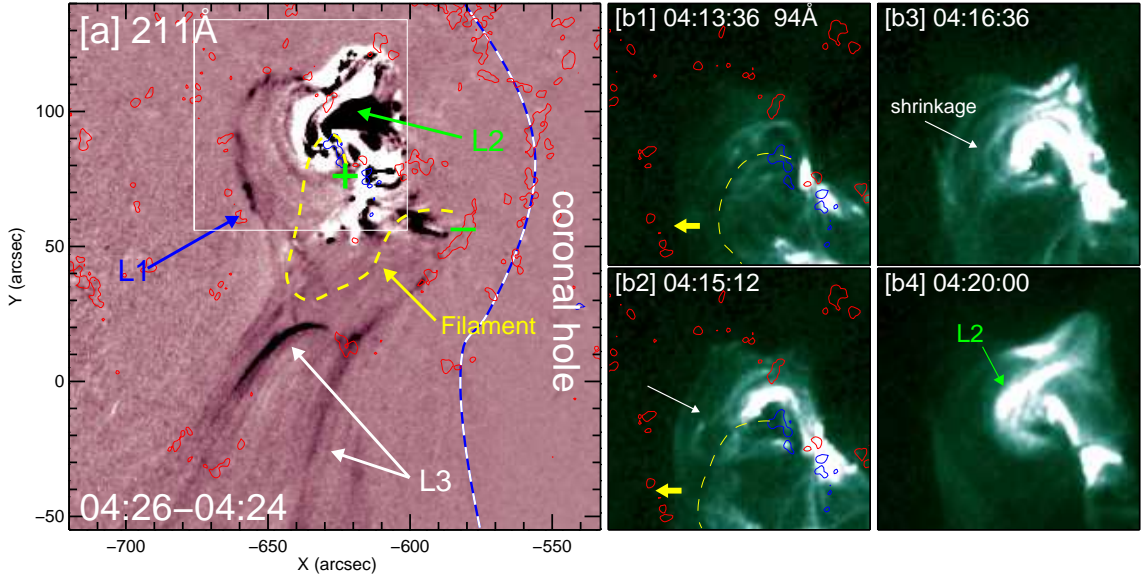}}
\caption{Panel (a): AIA 211 \AA\ running-difference image demonstrates the magnetic configuration, in which the yellow dashed line denotes of the filament; the green ``$\pm$'' symbols indicate the two feet of the filament; the blue arrow points to the background open field lines (L1); the green (L2) and white (L3) arrows show some newly formed loops; the white-blue dashed line marks the east outline of the nearby coronal hole; the white box denotes the FOV of (b1)--(b4). Panel (b1)--(b4): time sequence of AIA 94 \AA\ images show the formation process of L2. The yellow arrows indicate the moving direction of the filament which is indicated by the yellow dashed line; the green arrow denotes the shrinkage L2. The red and blue contours in panels (a), (b1) and (b2) are positive and negative magnetic field of the HMI LOS magnetogram scaled at ${\pm}$50 G, respectively. An animation of 211\AA\ and 94\AA\ is available. The animation covers 04:10-04:30 UT with 12 s cadence. The video duration is 3 seconds.}
\label{fig3}
\end{figure}

The erupting mini-filament transformed into a blowout jet via its magnetic reconnection with the ambient open magnetic fields. In \nfig{fig3} and its animation, the AIA multi-wavelengths observations clearly captured the magnetic reconnection process. To better illustrate the reconfiguration of the magnetic fields during this stage, we plot the changes of different loops in \nfig{fig3}. As shown in panel (a), the loop structures involved in the reconnection are denoted by arrows ''L1'', ''L2'', and ''L3'', respectively. Here, L1 refers to the pre-existing ambient open fields before the reconnection; L2 and L3 refer to the newly formed loops after the reconnection. By overlaying the axis of the erupting filament and the HMI LOS magnetogram on the AIA 211 \AA\ running-difference image at 04:26 UT (The running-difference image is obtained by subtracting the current image from the previous one that taken at 04:24 UT), one can easily identify that the north and the south feet of the filament were rooted in negative and positive magnetic regions, respectively. It is also noted that the ambient open field lines (L1) were rooted in a positive magnetic region. Therefore, it is clear that the directions of the magnetic fields of L1 and the axial field of the erupting filament were opposite, which is in favor of the occurrence of magnetic reconnection. Due to the interaction of the erupting filament with L1, the magnetic reconnection between them resulted in the generation of two groups of newly formed magnetic field lines, i.e., L2 and L3. L2 was observed as lower closed post-flare-loops, while L3 showed as open field lines and with the erupting filament mass along it (i.e., the jet). Moreover, the AIA 94 \AA\ images in panel (b1)--(b4) further displayed the rapid formation and shrinkage process of L2 (see the animation of \nfig{fig3}). As indicate by the yellow dashed curves, the erupting filament approached to the ambient open field lines and triggered the magnetic reconnection between them at about 04:13 UT (see panel (b1) and (b2)). Then, a group of closed and bright high temperature loop formed about one minute later (see panel (b3)), and it quickly shrunk down at about 04:20 UT. These remarkable topological changes well evidence the reconnection process that break the erupting filament and led to the formation of the blowout jet.

\subsection{Deflection of the Blowout Jet and the Formation of the CMEs}
\begin{figure}
\centerline{\includegraphics[width=0.8\textwidth,clip=]{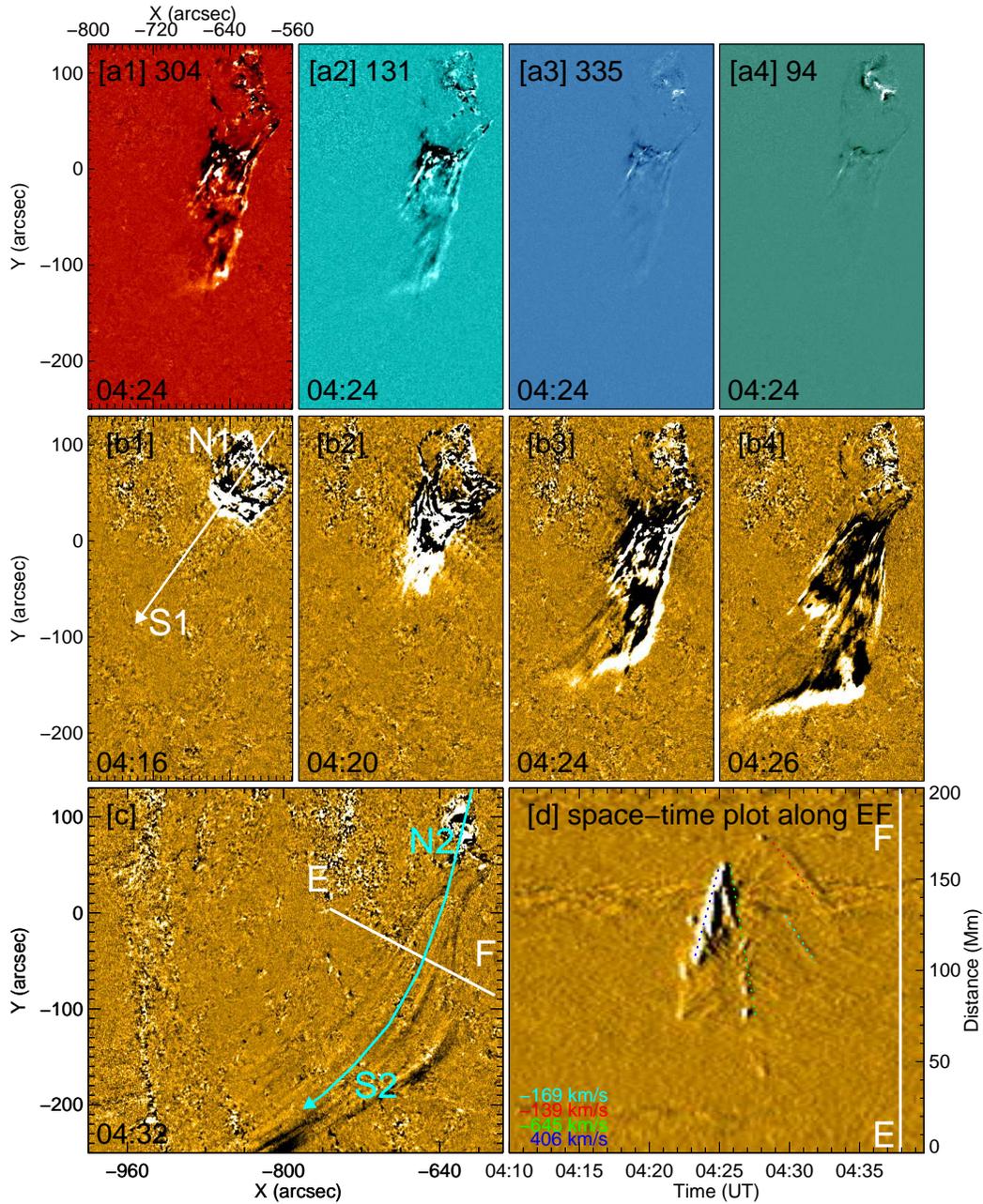}}
\caption{Panel (a1)--(a4): multi-wavelengths observation of the EUV jet. Panel (b1)--(b4): time sequence of AIA 171 \AA\ running-difference images demonstrate the eruption of the jet. In panel (b1) and (c), the white arrow N1-S1 and the cyan arrow N2-S2 mark the slit position of the time-distance plot shown in \nfig{fig7}(c1)--(c2) and (d1)--(d2), respectively. The time-distance plot along the white line E-F in panel (c) is shown in panel (d), which shows the expansion and contraction of the remote open coronal loops. The expansion and contraction motions of the loops are marked by four dashed lines in (d) with different colors. An animation of panels (c) is available. The animation covers 04:08-04:33 UT with 12 s cadence and the video duration is 5 seconds.}
\label{fig4}
\end{figure}

In this section, we focus on how did the blowout jet finally developed into a twin CME in the outer corona. The multi-wavelength images in \nfig{fig4}(a1)--(a4) show that the hot plasma component of the jet is not obvious in high-temperature channels, i.e, 94 and 335 channels). The time sequence AIA 171 \AA\ running-difference images in the middle row of \nfig{fig4} shows the ejection process of the blowout jet. As the filament reconnected with the ambient open fields, a spire structure formed ahead an U-shaped closed magnetic structure (see \nfig{fig4}(b1)). With the persistently break of the east leg of the U-shaped structure, the mass released from the closed filament became more conspicuous. Afterwards, filament plasma ejected along the newly formed open field lines (L3) and therefore formed the blowout jet. Consistent with previous observations, untwisting signals of the filament can also be discerned in the formation course of the blowout jet \citep[ see the animation of \nfig{fig4}]{2011ApJ...735L..43S,2012ApJ...745..164S,2019ApJ...873...22S}. Because of the releasing of the filament twist due to the magnetic reconnection between the filament and the ambient open fields, the blowout jet showed a wide spire and ejected outward with an obvious clockwise unwinding motion when looking above the jet axis.

It is striking that the ejection of the blowout jet was almost in the south direction at the beginning, then it changed to southeast direction at around 04:24 UT (see \nfig{fig4}(b3), (b4), and (c)). A time-distance along the line E-F as shown in \nfig{fig4}(c) is plotted in \nfig{fig4}(d), from which one can find that the jet body underwent an expansion and contraction motions during 04:22 UT to 04:33 UT. The speeds of the expansion and contraction motion of the jet body along the line E-F were about \speed{406} and \speed{169--645}, respectively.

\begin{figure}    
\centerline{\includegraphics[width=0.95\textwidth,clip=]{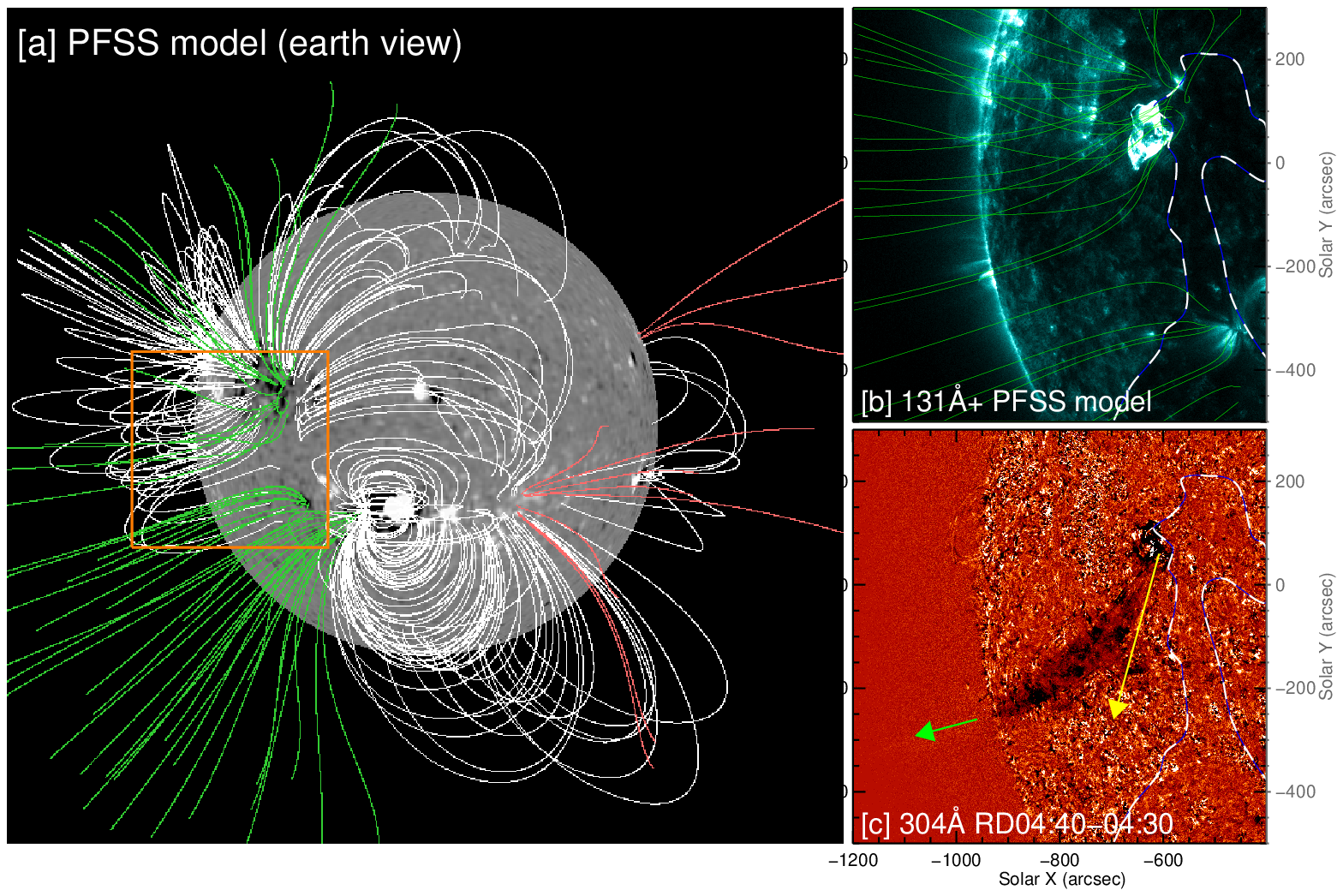}}
\caption{Panel (a): HMI full-disk magnetogram overlaid with the extrapolated magnetic field lines, in which the green and red lines show the open field lines originated from the negative and positive polarities, respectively. The brown box in (a) shows the FOV of panels (b) and (c). Panel (b): AIA 131 \AA\ images overlaid with the remote open field lines (green). Panel (c): AIA 304 \AA\ running difference image in which the yellow and green arrows indicate the eruption directions of the jet before and after the interaction to the remote open field lines. The white-blue dashed outlines the nearby coronal hole in (b) and (c).}
\label{fig5}
\end{figure}

\begin{figure}    
\centerline{\includegraphics[width=0.85\textwidth,clip=]{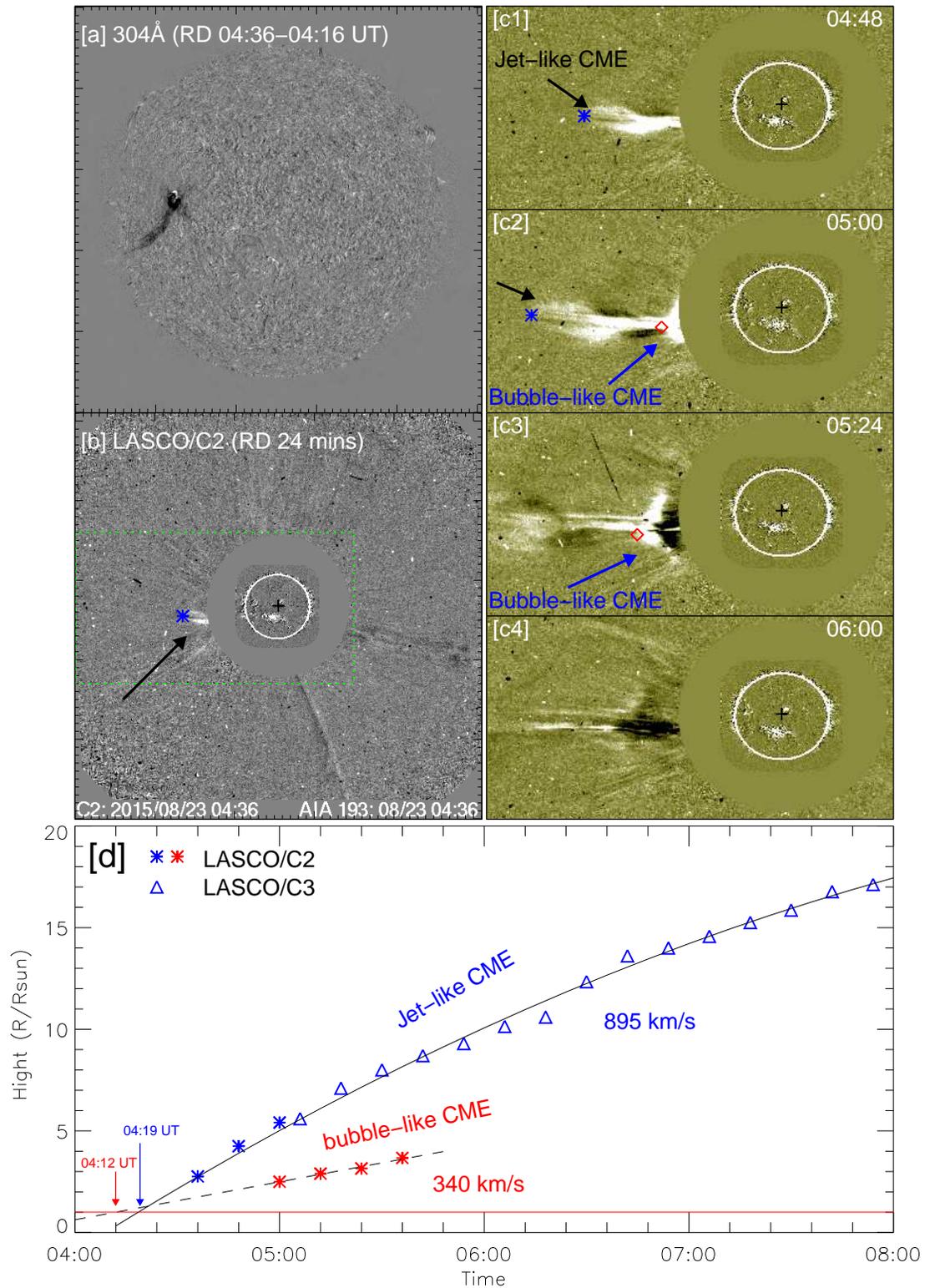}}
\caption{Panel (a) and (b): the running difference images of AIA 304\AA\ and LASCO/C2 images display the jet in inner and outer corona, respectively. The green dashed box denote the FOV of (c1)--(c4) that show the evolution of the twin CME. In panel (c1)--(c4), the blue asterisk and red diamond mark the front of the jet-like CME and bubble-like CME, respectively; The white circle and the black cross indicates the disk limb and the center of the Sun, respectively. Panel (d): the kinematics of the jet-like CME and the bubble-like CME. In which, the red line marks one solar radius. An animation of panel (b) is available. The animation covers 03:12-07:24 UT with 12 minutes cadence.}
\label{fig6}
\end{figure}

\begin{figure}[t]    
\centerline{\includegraphics[width=0.95\textwidth,clip=1]{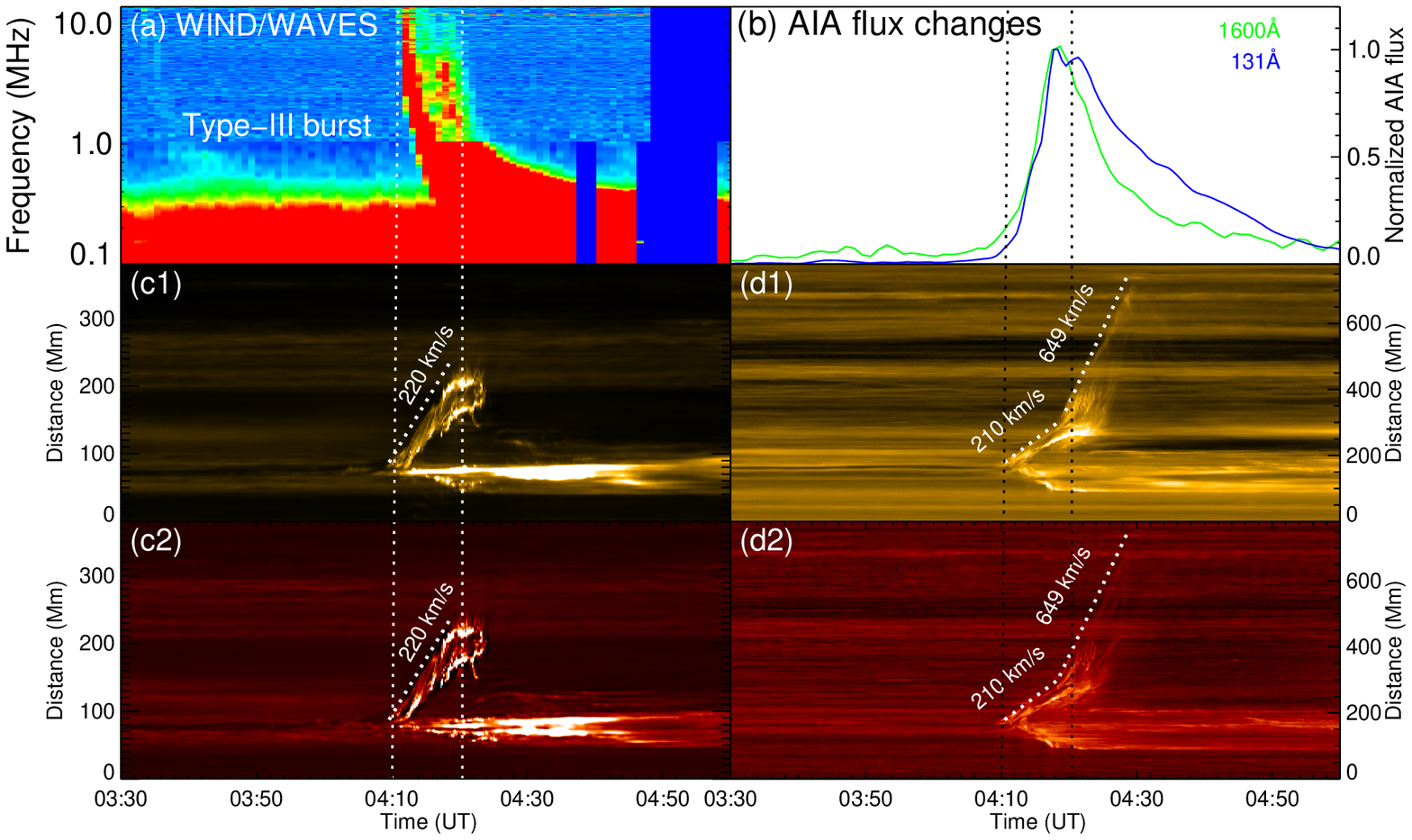}}
\caption{Panel (a): radio dynamic spectra obtained by RAD1 and RAD2 on board WIND/WAVES. Panel (b): AIA 1600 \AA\ (green) and 131 \AA\ (blue) lightcurves within the eruptive source region as outlined by the cyan dashed box in \nfig{fig2}(e). Panel (c1)--(c2): the time-distance plot along the line N1--S1 in \nfig{fig4}(b1). Panel (d1)--(d2): the time-distance plot along the line N2--S2 in \nfig{fig4}(c). The dotted black lines indicate the time (04:10 UT) and (04:20 UT), respectively.}
\label{fig7}
\end{figure}

The three-dimensional coronal magnetic field is extrapolated based on the photospheric magnetic field using the Potential Field Source Surface model \citep[PFSS][]{2003SoPh..212..165S} to investigate the physical reason that caused the direction change of the blowout jet (see \nfig{fig5}). The result of PFSS model can be obtained by using a software package in SolarSoftWare, which is based on synoptic magnetic maps from HMI with a 6 hour cadence. Some representative global field lines are shown in panel (a), in which the green and white lines are open and closed magnetic field lines, respectively. For better comparison, the AIA 131 \AA\ and 304 \AA\ running difference images of the brown box region shown in panel (a) are plotted in panel (b) and (c), respectively. The open field lines rooted in the eruption source region and the remote open field lines in the south hemisphere are overlaid in the AIA 131 \AA\ image, while the ejection directions of the blowout jet are indicated by the yellow and green arrows in the AIA 304 \AA\ running-difference image. It is clear that the change of the ejection direction of the blowout jet was caused by the blockage of another group of open field lines rooted in the south hemisphere, supporting the scenario that open magnetic flux can act as a potential wall so that solar eruptions can not penetrate it and have to be deflected away \citep{2009JGRA..114.0A22G,2009AnGeo..27.4491K,2018ApJ...862...86Y}.

Intriguingly, following the ejection of the blowout jet towards the outer corona, a twin CME, one was of jet-like shape and the other was of bubble-like shape, were captured by the {\em SOHO}/LASCO (see \nfig{fig6} and its animation). The jet-like CME was ahead of the bubble-like one. It is identified that the first appearance of the jet-like CME was at 04:36 UT in the FOV of LASCO/C2 (indicated by the black arrow in panel (b)), while that for the bubble-like CME was at 05:00 UT (indicated by the white arrow in panel (c2)). By comparison the initial eruption direction of the blowout jet in the inner corona (see panel (a)) and the jet-like CME in the outer corona (see panel (c1)), one can find that their eruption directions showed a large difference. As discussed in the above paragraph, the blowout jet was deflected by a group of remote open field lines in the south hemisphere. The eastward jet-like CME suggests that the angle between the injection and reflection directions of the blowout jet was about 90$^{\circ}$. The jet-like CME showed a collimated and elongated structure with an angular width of about ${\lesssim}15^{\circ}$, resembling many typical jet-like CMEs reported in previous studies \citep{1998ApJ...508..899W,2002ApJ...575..542W,2013SoPh..284..179V}. The bubble-like CME appeared 24 minutes later than the jet-like one in the FOV of LASCO/C2, and it has an angular width of about 48$^{\circ}$. The kinematic evolution of the jet-like CME and the bubble-like CME in LASCO are plotted in panel (d). By fitting the data  points with linear and second order functions, we obtain that the projection speed and deceleration of the jet-like CME were about \speed{895} and \accel{36}, respectively. The bubble-like CME can only be identified in the FOV of LASCO/C2 in four successful images; a linear fit to these data points yields that the projection speed of the bubble-like CME was about \speed{340}. Assuming the CMEs propagated with the obtained average speeds before they reached up to the FOV of LASCO/C2, we can extrapolate their onset times back to one solar radius when they were on the solar surface. As shown in panel (d), it is found that the onset time of the eruption of the bubble-like CME is around 04:12 UT (denote by the red arrow); while that for the jet-like CME is around 04:19 UT (denoted by the blue arrow). The extrapolated onset times of the jet-like and bubble-like CMEs are well agreement with the start times of the filament eruption and the generation of the blowout jet.

The kinematic evolution of the eruptive mini-filament/blowout jet, the lightcurves in the eruption source region, and the associated type-III radio burst are shown in \nfig{fig7}.
The kinematics of the mini-filament and the jet are shown in panel (c1)--(c2) and (d1)--(d2), respectively. Based on these time-distance diagrams, it is obtained that the speed of the mini-filament is about \speed{220}, while that for the jet is about \speed{649}. As shown in panel (b), the associated flare's start, peak, and end times were about 04:10 UT, 04:20 UT, and 04:30 UT, respectively. Moderate flux increase can be identified in the AIA 1600 \AA\ and 131 \AA\ lightcurves during the slow rising phase of the mini-filament (before 04:10 UT); then the violent eruption phase of the mini-filament corresponded the impulsive phase of the lightcurves. During this stage, the mini-filament erupted with the velocity of about \speed{210-220}. During the flare decay phase(after 04:20 UT), it can be seen that the jet had been accelerated to a speed of about \speed{649}. In addition, an obvious type-III radio bust was detected by the WIND/WAVES (see panel (a)), which rapidly drifted from about 10 MHz to a very low frequency ($\sim$ 0.1-0.3 MHz). Based on the leading radio burst theory, the appearance of type-III radio burst means  that a population of near-relativistic electrons stream from the eruptive source region to interplanetary space along open magnetic field lines \citep{2014RAA....14..773R,2017ApJ...851...67S}. Therefore, the type-III radio burst observed here evidenced the reconnection process between the rising filament and the ambient open field lines.

\section{Summary and Discussion} \label{sec:summ}
In this paper, we present a direct imaging observation of a twin CME evolved from a coronal blowout jet that involved the eruption of a mini-filament in the jet's eruption source region. The generation of the blowout jet was due to the magnetic reconnection between the rising filament and the ambient open coronal loops. It is interesting that the ejection of the blowout jet experienced an obvious deflection of about $90^{\circ}$ due to the blockage of a group of remote open coronal field lines in the south hemisphere. The deflection significantly changed the ejection direction of the blowout jet from southward to almost eastward. Analysis results of the photospheric magnetic fields suggest that the activation and slow rising of the mini-filament was associated with the magnetic flux cancellation in the filament channel, while the sudden emergence of the positive magnetic flux was possibly responsible for the violent fast eruption of the filament. Moreover, quasi-periodic intermittent jet-like activities are also observed in the filament channel in the course of flux cancellation, which may also play a role in the filament eruption \citep[see also,][]{2018NewA...65....7T}. Comparing to previous studies, the present event suffered less projection effect, and it can provide more accurate information for mini-filament driven blowout jets, as well as the twin CME.

CMEs have been studied for many years, and various models are proposed to interpret their formation, evolution, and propagation mechanisms \citep{2011LRSP....8....1C}. The most previous observations indicated that a single solar eruption in the inner corona always results in a corresponding large-scale CME. Only a small fraction of solar eruptions can launch multiple CMEs in different ways. For example, \cite{2008ApJ...677..699J} reported that a sympathetic CME pair was physically connected to an on-disk jet, in which one was related to the flare and jet, while the other was due to the eruption of a interconnecting loop caused by the interaction of the jet plasma. \cite{2012ApJ...750...12S} reported the sympathetic filament eruptions in a breakout magnetic configuration; they found that the two adjacent filaments erupted sequentially and proposed that there would be two corresponding CMEs in the outer corona within the framework of the magnetic breakout model. Global EUV waves can propagate to a large distance and can trigger multiple solar eruptions in their path \citep{2014ApJ...786..151S,2014ApJ...795..130S}; therefore, they could potentially result in the possible occurrence of multiple solar eruptions and CMEs within a short time interval.

Nevertheless, so far reports on simultaneous appearance of a pair of CMEs evolved from one single jet are still very scarce in literature. The first report on such kind of eruption was published by \cite{2012ApJ...745..164S}, where they found a typical filament driven blowout jet which showed cool dark and hot bright components in the inner corona, whereas in the outer corona they detected a pair of simultaneous CMEs, i.e., a jet-like CME and a bubble-like CME. The authors also proposed a model to interpret this interesting eruption, in which the jet-like CME is first resulted from the hot bright jet component that produced by a preceding external reconnection between the rising confining fields of the mini-filament and the ambient open field lines, while the bubble-like one is subsequently developed from the ensuing eruption of the mini-filament due to the removal of the confining field of the mini-filament by the external reconnection.

In the present case, a similar twin CME was also observed by LASCO after the mini-filament eruption. Combining the STEREO-A EUVI and SDO/AIA observations, we found that there is no any other obvious eruptions occurred from 03:30 UT to 05:00 UT, which allowed us exclude the backside eruptions. In order to relate the twin CME back into their solar counterparts, we further investigate the kinematics of the jet-like and the bubble-like CMEs, respectively. The height-time relationship of the twin CME have been presented in \nfig{fig6}(d). In which, one can clear see that the jet-like one first appeared in the FOV of LASCO/C2 at around 04:36 UT, and finally propagated to 17 solar radius at an averaged speed of around \speed{895}; while the bubble-like CME had only been captured by LASCO/C2 during 05:00-05:36 UT, which had a speed of about \speed{340} and faded away within 5 solar radius. The extrapolated onset times of the bubble-like and the jet-like CMEs are about 04:12 UT and 04:19 UT, respectively. Based on the close temporal and spatial relationship between the twin CME and their solar counterparts, we conclude that the bubble-like CME should correspond to the eruption of the mini-filament that started at around 04:10 UT and with the a speed of about \speed{210-220}, while the jet-like CME should be the coronal extension of the blowout jet that began at about 04:20 UT and with a speed of about \speed{649}. It is worthy to note that in our event the loss-of-equilibrium of the mini-filament happened at first, while the external reconnection should be triggered due to the interaction between the rising filament and the background open field, which is different form the event reported by \citep{2012ApJ...745..164S}. Although we did not detect this bright hot jet component during the eruption in the inner corona (see the \nfig{fig4}(a1)--(a4)), the detected type-III radio burst and the rearrangement of the coronal loops do indicate the occurrence of the reconnection.

Our analysis results indicate that the start time of the jet-like CME was after the bubble-like one for a few minutes; however, the former was captured by LASCO earlier than the later, which implies that the jet-like CME underwent a more prominent acceleration after its initiation. The ejecting speed of the jet-like CME was about \speed{895}, such a high speed was possibly due to the following reasons. We note that the jet experienced an interaction process with a group of remote open field lines that caused the deflection of the ejecting direction of the jet plasma. During the interaction process, we do not find any signature of energy dissipation at the interaction site. The significant acceleration of the jet plasma may mainly due to the acceleration of the slingshot effect of the bended open loops cased by the interaction. The untwisting motion of the jet represent the releasing of non-potential magnetic energy stored in the pre-eruption filament, which can also contribute to the acceleration of the jet \citep{2011ApJ...735L..43S}. In addition, we argue that during the interaction period, a part of jet plasma may stay at the interaction site or fall back to the solar surface; therefore, only a portion of initial ejecting plasma was accelerate to the outer corona. In that case, the total kinetic energy of the initial jet plasma was transformed into the kinetic energy of the portion of the ejected plasma after the interaction. Due to the decreased mass, the ejecting portion of the jet plasma can also obtain a higher speed. Regarding to bubble-like CME, its speed was about \speed{340}. Such a speed indicate that the bubble-like CME should be a slow CME that is often associated with filament eruption \citep{gosling76}. In the present event, the bubble-like CME did associated with the eruption of a mini-filament, which can possibly account for its relatively lower ejecting speed.

Many observations have revealed that solar eruptions can be channeled and guided by neighboring large-scale background magnetic fields during their initial stage. When erupting magnetized plasma interact with remote magnetic structures, they are often deflected to the direction of minimum magnetic force on global scales \citep{2015A&A...583A.127N}. In the highly complicated corona, there are many magnetic structures (i.e., active regions, coronal holes, and helmet streamers) can act as obstacles and impose actions on solar eruptions impeding on them. Previous studies suggested that erupting filaments may tend to propagate to the region with weak field or lower magnetic energy density \citep{2011NewA...16..276B,2011SoPh..269..389S}, CMEs can be channeled and guided by large-scale magnetic fields such as streamers and open fields in coronal holes \citep{2001JASTP..63..389P,2001SoPh..203..119F,2005ApJ...635L.189B,2007ApJ...667L.105J,2013ApJ...773..162B,2016ApJ...819L..18Z,2018ApJ...862...86Y}. For the interaction between CMEs and open field lines such as in coronal holes, \citet{2009JGRA..114.0A22G} proposed two possible consequences for the interaction depending on their relative directions of the field lines. When the magnetic fields of the CME are parallel to the open field lines, the CME will be deflected due to its lower horizontal speed component relative to the Alfv\'en speed in the coronal hole. When the magnetic field lines are antiparallel to each other, magnetic reconnection can be expected between them.  Aided by the PFSS model, it is identified that the deflection of the blowout jet or CME in the present study was due to the blockage of a group of remote open field lines rooted in the south hemisphere. The extrapolated coronal field lines reveals that the magnetic field of the blowout jet and the remote open field lines in the south hemisphere were parallel, therefore, there should be no reconnection between them. In addition, the extrapolated coronal magnetic field also revealed that the eruption direction of the CMEs were consistent with that of the bent open field lines. Therefore, these field lines also channeled the propagation of the CMEs.

\acknowledgments
The authors would like to thank the data provided by the {\em SDO}, {\em SOHO}/LASCO, {\em WIND} science teams, the anonymous referee's valuable comments and suggestions that largely improved the quality of the paper, and the helpful discussions with Dr. Yu Liu and Junchao Hong from Yunnan Observatories, CAS, China. This work is supported by the Natural Science Foundation of China (11773068, 1163308, 11363007), the Yunnan Science Foundation (2017FB006), and the West Light Foundation of Chinese Academy of Sciences. H.C.C. acknowledges the support by NSFC 11703084, 11633008, 11873088.



\end{document}